\documentclass[preprint2]{aastex}

\usepackage{natbib}

\slugcomment{Version 13/04/25}

\shorttitle{Magnetically Confined Hot Plasma\\
 in the Nuclear Bulge}
\shortauthors{Nishiyama et al.}

\begin{document}

\title{
Magnetically Confined Interstellar Hot Plasma\\
in the Nuclear Bulge of our Galaxy}

\author{Shogo Nishiyama\altaffilmark{1},
Kazuki Yasui\altaffilmark{2}, 
Tetsuya Nagata\altaffilmark{2},
Tatsuhito Yoshikawa\altaffilmark{2},
Hideki Uchiyama\altaffilmark{3},
Rainer Sch\"{o}del\altaffilmark{4},
Hirofumi Hatano\altaffilmark{5},
Shuji Sato\altaffilmark{5},
Koji Sugitani\altaffilmark{6},
Takuya Suenaga\altaffilmark{7},
Jungmi Kwon\altaffilmark{1},
and Motohide Tamura\altaffilmark{1}
}

\altaffiltext{1}{National Astronomical Observatory of Japan, 
Mitaka, Tokyo 181-8588, Japan}

\altaffiltext{2}{Department of Astronomy, Kyoto University, 
Kyoto 606-8502, Japan}

\altaffiltext{3}{Department of Physics, The University of Tokyo, 
Bunkyo-ku, Tokyo 113-0033, Japan}

\altaffiltext{4}{Instituto de Astrof\'{i}sica de Andaluc\'{i}a (IAA)-CSIC,
18008 Granada, Spain}

\altaffiltext{5}{Department of Astrophysics, Nagoya University, 
Nagoya 464-8602, Japan}

\altaffiltext{6}{Graduate School of Natural Sciences, Nagoya City University,
Nagoya 467-8501, Japan}

\altaffiltext{7}{Department of Astronomical Sciences, Graduate University for Advanced Studies (Sokendai),
Mitaka, Tokyo 181-8588, Japan}


\begin{abstract}

The origin of the Galactic center diffuse X-ray emission (GCDX)
is still under intense investigation.
In particular, the interpretation of 
the hot ($kT \approx 7$\,keV) component of the GCDX,
characterised by the strong Fe 6.7\,keV line emission,
has been contentious.
If the hot component originates from a truly diffuse interstellar plasma,
not a collection of unresolved point sources,
such plasma cannot be gravitationally bound,
and its regeneration would require a huge amount of energy.
Here we show that the spatial distribution of the GCDX 
does {\it not} correlate with the number density distribution 
of an old stellar population traced by near-infrared light,
strongly suggesting a significant contribution 
of the diffuse interstellar plasma.
Contributions of the old stellar population to the GCDX
are implied to be $\sim 50$\,\% and $\sim 20$\,\% 
in the Nuclear stellar disk and Nuclear star cluster, respectively.
For the Nuclear stellar disk, 
a scale height of $0\fdg32 \pm 0\fdg02$ is obtained for the first time
from the stellar number density profiles.
We also show the results of the extended near-infrared polarimetric observations 
in the central $3\degr \times 2\degr$ region of our Galaxy,
and confirm that the GCDX region is permeated by 
a large scale, {\it toroidal} magnetic field as previously claimed.
Together with observed magnetic field strengths close to energy equipartition,
the hot plasma could be magnetically confined,
reducing the amount of energy required to sustain it.

\end{abstract}

\keywords{Galaxy: center --- X-rays: ISM ---  polarization --- ISM: magnetic fields}


\section{INTRODUCTION}

In the late 80s, a Japanese X-ray satellite {\it GINGA}
revealed the presence of a diffuse and rather uniform 
6.7\,keV emission from highly ionized, Helium-like ions of iron 
at the direction of the Galactic center \citep[GC; ][]{89KoyamaNat}.
The line emission and associated continuum component,
called the Galactic center diffuse X-ray emission (GCDX),
resembles the Galactic ridge diffuse X-ray emission 
\citep[GRXE; e.g.,][]{Cooke69,86KoyamaGRX}
extending more than $100\degr$ along the Galactic plane.
For the GRXE, 
more than $80\,\%$ of the diffuse emission 
has been claimed to be resolved
into point sources \citep{09RevnivtsevNat},
suggesting faint X-ray point sources in origin.

On the other hand, the origin of the GCDX,
in particular its very hot component 
with a temperature of $kT \sim 7$\,keV,
is more puzzling.
So far only 10 - 40\,\% of the GCDX has been claimed to be resolved
into faint point sources even with the {\it Chandra} satellite
\citep{Muno04DifXGC,07RevnivtsevGRXB}.
The plasma temperature, represented by flux ratios of iron emission lines,
is systematically higher for the GCDX than for the GRXE \citep{Yamauchi09},
indicating their different origins.

Two main ideas have been suggested to account for it :
a truly diffuse plasma that bathes the emitting region 
\citep[e.g.,][]{89KoyamaNat};
and a superposition of a large number of unresolved point sources
as the GRXE is \citep[e.g.,][]{Wang02Nat}.
In the later case, candidates are old stellar binary systems such as 
cataclysmic variables (CVs) 
and coronally active binaries \citep[ABs;][]{06Sazonov}.
So if the hot component originates in the discrete sources,
its spatial distribution should be very similar to 
that derived by old stars observable in infrared wavelengths.
For this purpose, 
a stellar mass distribution model constructed from 
infrared surface brightness maps \citep{Launhardt02} has been used
\citep{Muno09,11Uchiyama,Heard13}. 
However, such maps could be subject to the influence of bright stars.
The angular resolution in these maps was only 0\fdg7,
so that the stellar density profile in the direction
orthogonal to the Galactic plane had to be inferred from proxies
(dust emission, radio emission from molecular clouds).
In addition, uncertainties of the mass model seem to be
as high as a factor of two \citep{Launhardt02}.

We have constructed a stellar {\it number density} map
of the GC region from new near-infrared (NIR) observations 
with more than 1,000 times higher spatial resolution (Yasui et al. in preparation),
which enables us to directly compare the stellar distribution with GCDX.
In this {\it letter}, 
we summarize the NIR imaging observations and their results,
and provide additional evidence for the hypothesis
that the GCDX arises from a truly diffuse hot plasma.
We also show results of our recent polarimetric observations. 
The results provide strong evidence for a large-scale
toroidal magnetic field configuration
which could confine the hot plasma magnetically.

\section{Observations and Data Analysis}
\label{sec:obs}

The central region of our Galaxy,
$\mid l \mid \la 3\fdg0$ and $\mid b \mid \la 1\fdg0$
(corresponding to 840\,pc $\times$ 280\,pc at 8\,kpc from the Sun),
was observed from 2002 to 2004
using the NIR camera SIRIUS \citep{99Nagashima,03NagayamaSPIE}
on the 1.4 m telescope IRSF.
SIRIUS provides $J$ (1.25 $\mu$m), $H$ (1.63 $\mu$m),
and $K_S$ (2.14 $\mu$m) images simultaneously.
The averages of the 10$\sigma$ limiting magnitudes are 
$H=16.6$ and $K_S=15.6$.
We do not use the $J$-band data due to severe interstellar extinction.
Further details are given in \citet{Nishi06Ext}.

The stellar number density map is constructed as follows:
At first, an $H$ and $K_S$ color magnitude diagram (CMD) is constructed
for each sub-field of $20\arcmin \times 20\arcmin$,
and foreground sources with their blue $H-K_S$ color are removed.
The typical color of stars in the GC is $H-K_S > 1.0$,
and color cuts to remove the foreground sources are 0.3 - 1.1.
We carry out an extinction correction for each star using 
the observed $H-K_S$ color,
the mean intrinsic color of $(H-K_S)_0 \approx 0.20$
\citep[considering the limiting magnitudes and the Galactic model by][]{Wainscoat92},
and an interstellar extinction law, 
$A(K_S) = 1.44 \times \{(H-K_S) - (H-K_S)_0\}$ \citep{Nishi06Ext}.
Here we obtain an extinction-corrected $K_S$-band magnitude,
$K_{S,0}$,
and the amount of interstellar extinction for each star.

\begin{figure}[h]
 \begin{center}
   \epsscale{1.10}
   \plotone{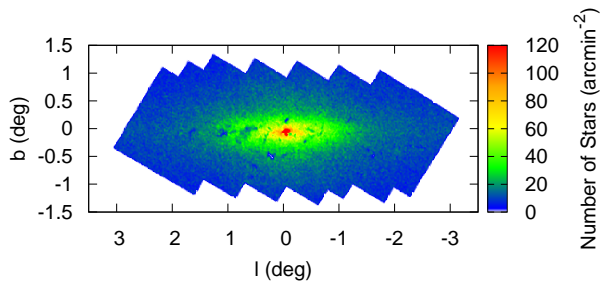}
 \end{center}
    \caption{
      The stellar number density map of 
      the central $6\degr \times 2\degr$ region of our Galaxy. 
      Stars with $K_{S,0} < 8.0$ and $< 10.5$
      are used for the central $20\arcmin$,
      and for the outside of $20\arcmin$, respectively. 
      Several low-density regions are seen in the NB,
      and are not used in the following analysis.
    }
    \label{fig:DensityMap}
\end{figure}

We then construct a stellar number density map
using stars with $K_{S,0} <10.5$ (Fig. \ref{fig:DensityMap}).
For the central $20\arcmin \times 20\arcmin$ field, 
source confusion is so severe that
a different magnitude limit of $K_{S,0} < 8.0$, 
and a conversion factor derived by the ratio 
of the number of stars with $K_{S,0} < 8.0$ and $K_{S,0} < 10.5$
are used.
By comparing $K_S$-band luminosity functions constructed with 
the extinction-corrected stars and 
a Galactic model \citep{Wainscoat92},
these magnitude limits are determined so that
completeness at the limits is almost 100 \%.
We make completeness corrections with recovery rates determined 
by adding artificial $K_S=12.5$ sources in the images, 
and confirm that the recovery rates are very high, 
$\sim 96$\,\% in average \citep{Hatano13}.
For more detail, see Yasui et al. (in preparation).
Also, at the very center ($< 1\arcmin$),
we have used images obtained with an 8-m telescope VLT
and ISAAC \citep{Nishi13NSC}.

To determine a large-scale interstellar MF configuration, 
we have carried out NIR polarimetric observations
using IRSF and 
a NIR polarimetric imager SIRPOL \citep{Kandori06}, from 2006 to 2010.
We have extended the survey region from 
$\mid l \mid \la 1\fdg0$ 
\citep[and $\mid b \mid \la 1\fdg0$;][]{Nishi10}
to $\mid l \mid \la 1\fdg5$,
which covers almost the whole region dominated by GCDX.
Comparing the polarization between stars distributed
further and closer side in the GC,
we obtain polarization originating from
magnetically aligned dust grains in the GC
\citep[for more detail, see][]{Nishi09MF,Nishi10}.
The polarized angle traces the GC's MF direction projected onto the sky.

\section{Results and Discussion}

\subsection{Stellar Number Density Profiles}

\begin{figure}
 \begin{center}
   \epsscale{.90}
    \plotone{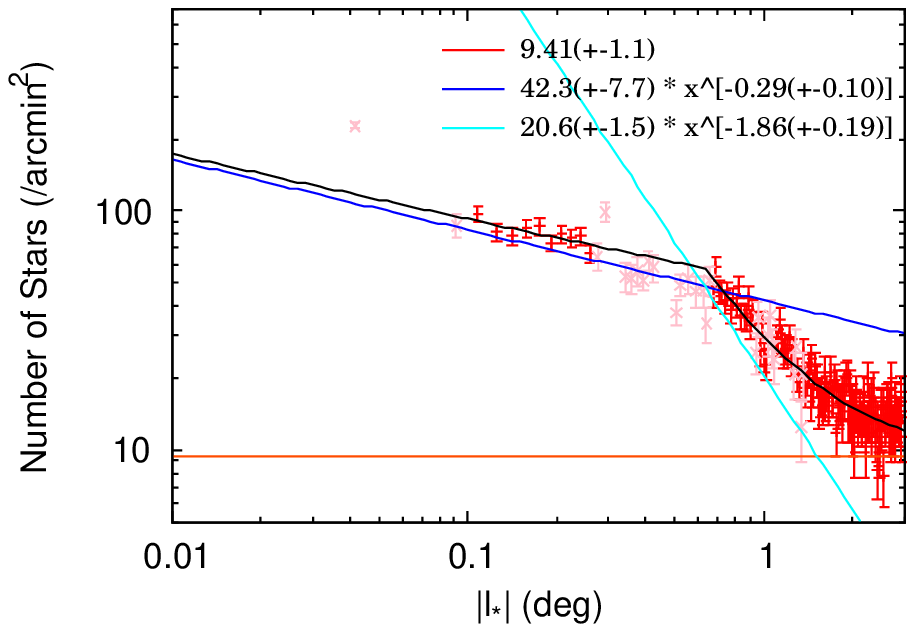}
    \plotone{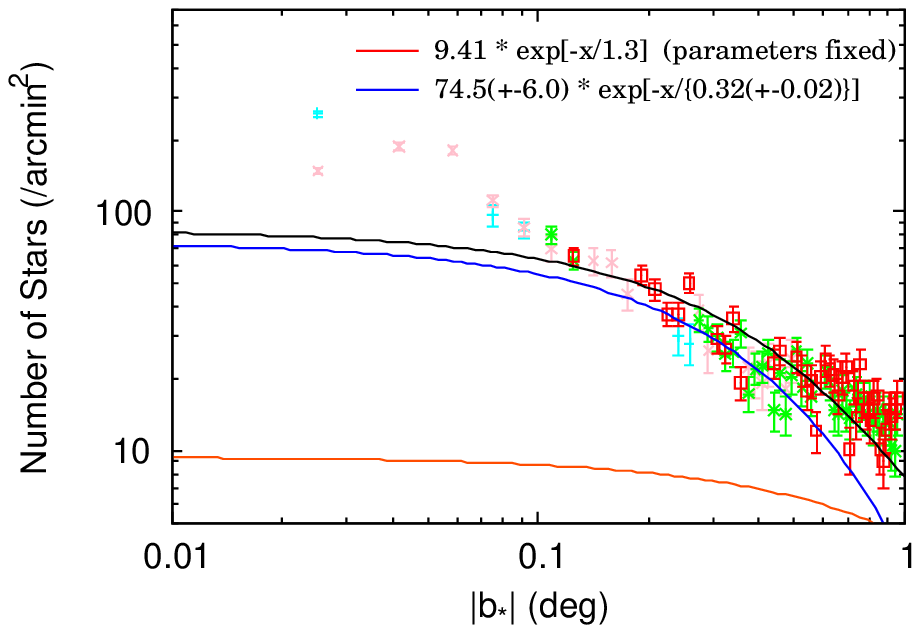}
 \end{center}
    \caption{
      Top: Longitudinal profile
      of the stellar density distribution after a completeness correction. 
      Only data points at $l_* < 0$ is plotted,
      and those with pink marks are not used for the fitting.
      The profile can be approximated 
      by a sum (black line) of
      two different power laws (blue and cyan lines) 
      and a constant component (red line).
      The boundary of the two power laws,
      $l_* = 0\fdg63$,
      corresponds to the outer edge of the inner NB \citep{Launhardt02}.
      Bottom:
      Latitudinal stellar density profiles.
      Data points from positive and negative $b_*$ 
      are plotted as green (cyan) and red (pink) marks, respectively.
      The cyan and pink marks are not used for the fitting
      because the number densities are underestimated
      in the corresponding regions due to strong line-of-sight extinction.
      Data points at $\mid b_* \mid < 0\fdg1$ are also not used for fitting
      because of the dominance of the NSC.
      The profile is fit with the NSD (blue line) and 
      the Galactic disk (GD; red line) components
      using exponential functions, 
      where the scale height of the GD is assumed
      to be  $1\fdg3$.
      The NSD component has a scale height of $0\fdg32 \pm 0\fdg02$.
      The number densities are calculated in rectangular bins
      with a size of $1\arcmin (l) \times 2\arcmin (b)$
      for the longitudinal profile at $b_* = 0\degr$,
      and $2\arcmin (l) \times 1\arcmin (b)$
      for the latitudinal profile at $l_* = 0\degr$.
      }
   \label{fig:ProfileFit}
\end{figure}

A disk-like structure is seen 
in the stellar number density map (Fig. \ref{fig:DensityMap}).
This is known as the Nuclear bulge (NB),
which consists of the Nuclear stellar disk (NSD)
and the Nuclear stellar cluster \citep[NSC;][]{Launhardt02}.
Here our observations clearly reveal morphology of the NB on a large scale,
with much higher spatial resolution
than previous studies.
The NB has a symmetric, disk-like structure
with a scale height of 
$0\fdg32 \pm 0\fdg02$ (Fig. \ref{fig:ProfileFit}),
although several low-density regions are seen.
Those are very dense molecular clouds 
in front of/inside the NB.
A true stellar number density is difficult to be derived in these regions,
and thus they are not used in the following analysis (see Yasui et al. in preparation).

\begin{figure}
   \epsscale{1.15}
   \plotone{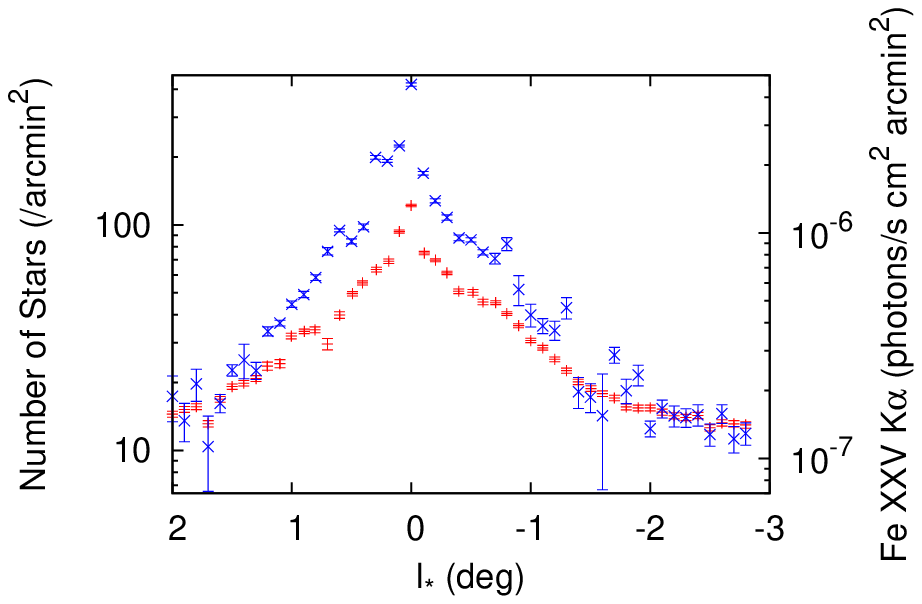}
   \plotone{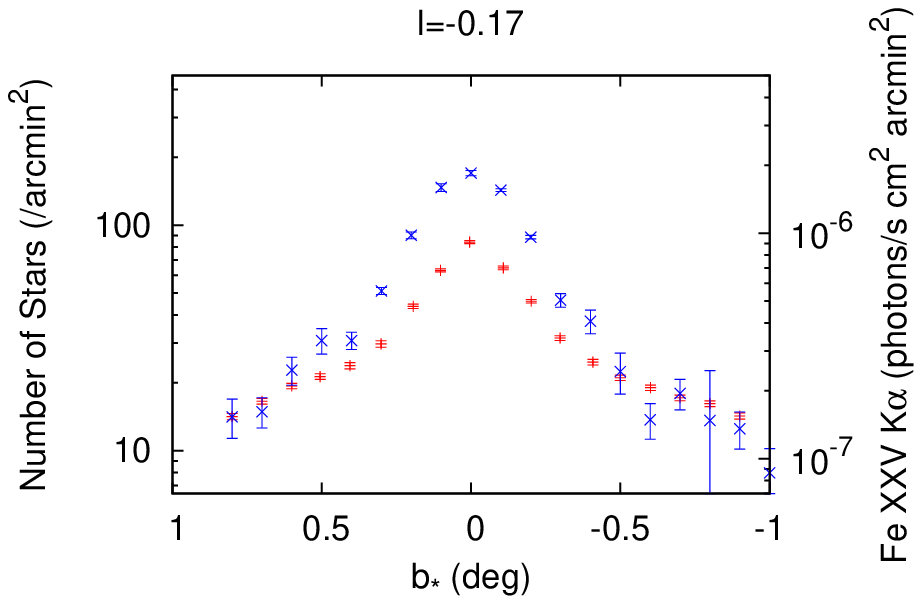}
    \caption{
      Longitudinal (top) and latitudinal (bottom) profiles
      of the stellar number density after a completeness correction 
      (red crosses).
      Overplotted are the 6.7\,keV-emission profiles
      \citep[blue x;][]{07KoyamaFe,11Uchiyama}.
      The region outside the NB, 
      $1\fdg5 \leq \mid l_* \mid \leq 2\fdg8$,
      is used to scale the the 6.7\,keV emission profile
      to have the same value as the stellar number density.
      The number density is calculated in the same rectangles
      as those used in \citet{11Uchiyama},
      with a size of $0\fdg1 (l) \times 0\fdg2 (b)$
      for the longitudinal profile,
      and $0\fdg2 (l) \times 0\fdg1 (b)$
      at the position of $l = -0\fdg17$
      for the latitudinal profile.
      The same scaling factor is used for the latitudinal profile.
    }
   \label{fig:DensityProfile}
\end{figure}

The longitudinal and latitudinal profiles
of the 6.7\,keV line emission measured by {\it Suzaku} 
\citep{07KoyamaFe,11Uchiyama}
clearly show an excess at the NB region
over the stellar number density profiles 
(Fig. \ref{fig:DensityProfile}).
The two profiles are overplotted and scaled
to have the same values at $1\fdg5 < \mid l_* \mid < 2\fdg8$,
i.e., in a region outside of the NB
[$l_*$ and $b_*$ denote the angular distance from Sgr\,A* 
along the Galactic longitude and latitude, respectively,
and $(l_*, b_*) = (l + 0\fdg056, b + 0\fdg046)$].
Fitting the longitudinal profile in the range $-0\fdg7 \leq l_* \leq -0\fdg1 $
with a power-law of $\propto \theta^{-\alpha}$,
where $\theta$ is angular offset from Sgr A*, gives 
$\alpha_{\mathrm{star}} = 0.30 \pm 0.03$ 
for the stellar number density. 
This is different both from $0.44 \pm 0.02$ 
for the 6.7\,keV profile in the same range,
and from $0.60^{+0.02}_{-0.03}$ for the integrated emission
of Fe\,6.7 and 6.9\,keV lines \citep{Heard13}.

The majority of faint X-ray 
($L_{\mathrm{2-10\,keV}} < 10^{30}$\,erg\,s$^{-1}$) sources 
which have not been resolved
but contribute to the GCDX are most likely to be 
old binary systems \citep{06Sazonov}.
Using the synthetic CMD computation \citep{Aparicio04},
and a constant star formation history 
during 13\,Gyr
for the NSD \citep{Figer04SFH},
we have confirmed that about 75\,\% of the stars
with $K_{S,0} < 10.5$ are older than 1 Gyr.
So the NIR map and profiles shown here trace 
the distribution of the old stars,
and they are clearly different from those of the 6.7\,keV emission.

\begin{figure}
 \begin{center}
   \epsscale{1.2}{
   \plotone{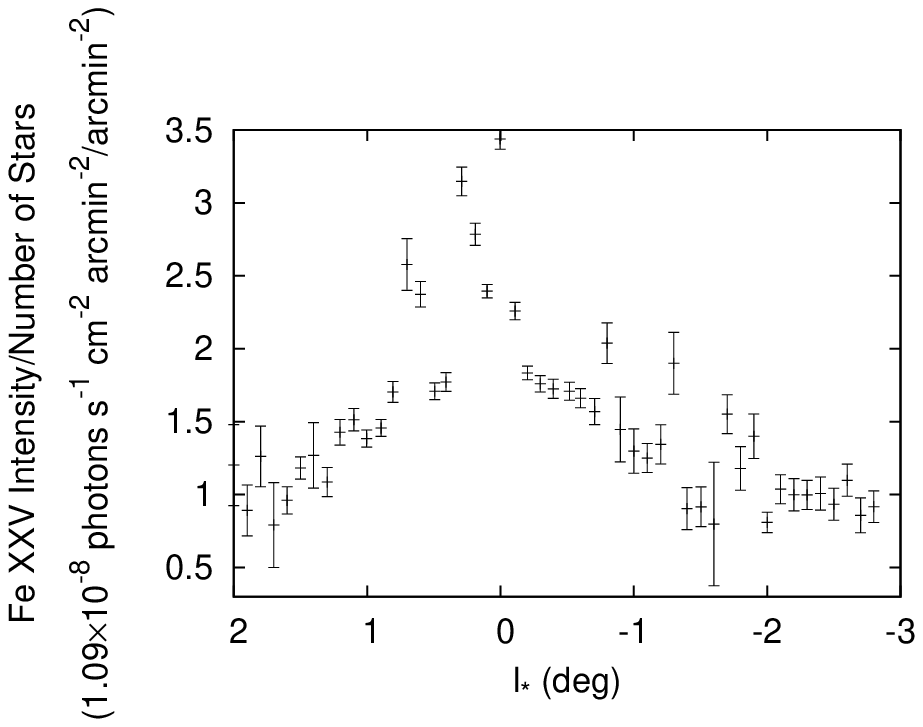}
   \plotone{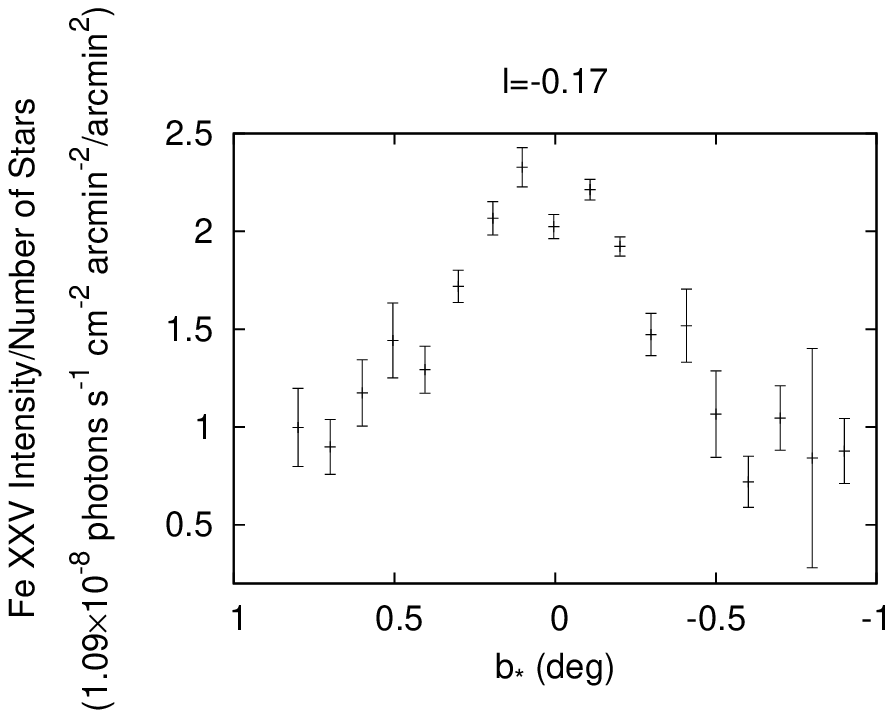}
 }
 \end{center}
    \caption{
      Longitudinal (top) and latitudinal (bottom) profiles
      of the ratios of the 6.7\,keV emission to the stellar number density,
      scaled to be unity at the position for normalization,
      $1\fdg5 \leq \mid l_* \mid \leq 2\fdg8$.
      These profiles represent a contribution of point sources,
      traced by our NIR observations, to the GCDX
      in the assumption that the contribution of truly diffuse hot plasma
      is negligible at the position for normalization 
      (i.e., the Galactic ridge region).
    }
   \label{fig:DensityProfile2}
\end{figure}

A contribution from faint discrete sources to the GCDX has been claimed
\citep{Wang02Nat,Muno04DifXGC,07RevnivtsevGRXB}.
To investigate the contribution of the point sources,
especially of the old stellar population detectable in our observations,
we construct longitudinal and latitudinal profiles for the ratio
of the 6.7\,keV emission to the stellar number density
(Fig. \ref{fig:DensityProfile2}).
When the profiles are scaled to be unity at 
$1\fdg5 < \mid l_* \mid < 2\fdg8$,
the ratios are $\sim 1.5$ and $\sim 3$ in the NSD and NSC, respectively.

The NB and NSC have a different formation history from the 
Galactic bulge (GB),
and have formed stars over their entire lifetime,
indicating more bright stars in the NB and NSC.
Here we use the synthetic CMD computation \citep{Aparicio04}
to estimate the fraction of 
the total number of stars formed in the computation, $N_{\mathrm{all}}$,
to the number of stars with $K < 10.5$, $N_{K<10.5}$.
This ratio, $R \equiv N_{\mathrm{all}}/N_{K<10.5}$,
represents the ratio of the theoretically expected total number of stars
to the number of stars detected in our observations.
The star formation histories used here are:
a burst star formation from 10 to 13\,Gyr ago 
for the GB \citep[i.e., outside the NB;][]{Zoccali03};
a constant star formation rate for 13\,Gyr for the NSD \citep{Figer04SFH};
and the history derived by \citet{Pfuhl11} for the NSC.
When we scale the ratio $R$ to be unity for the GB,
we obtain $R_{\mathrm{GB}} : R_{\mathrm{NSD}} : R_{\mathrm{NSC}} \approx 1 : 0.8 : 0.6$.
This result means that the densities of the old stellar population
in the NSD and NSC are over-predicted by the bright stars.
Taking into account this ratio, we have found that
the contributions of the old stellar population to the GCDX are
$(1/1.5) \times 0.8 \sim 0.5$ and $(1/3) \times 0.6 \sim 0.2$ 
for the NSD and NSC, respectively,
in the assumption that
the X-ray luminosity function is universal,
and that the contribution of point sources
to the GRXE is 100\,\%.
The contribution at the NSC is in good agreement with $\sim 1/6$
derived by \citet{Koyama09}.

A larger X-ray emissivity per unit stellar mass 
for the GCDX than the GRXE has been claimed
to explain the different distributions of stars and 6.7\,keV emission
\citep{07RevnivtsevGRXB,Heard13}.
To change the emissivity, at least one of 
initial mass function (IMF), 
binary fraction (BF), or
star formation history (SFH)
is required to be different in the NB from the GB.
In the preceding paragraphs, we have shown that 
different SFHs cannot explain the different spatial distributions
of the stellar number density and 6.7\,keV emission.
Considering an universal IMF \citep{Bastian10ARAA}
and a possible top-heavy IMF in the GC \citep{Figer99},
the number of old, low-mass stars (i.e., CVs and ABs) per unit stellar mass
never increases, it only decreases.
Also, a higher stellar density tends to destroy binaries 
rather than to form them via a capture process,
which seems to play a small role in binary formation \citep{Tohline02}.
These imply a smaller X-ray emissivity per unit stellar mass 
{\it by CVs and ABs}, rather than a higher emissivity.

\subsection{Magnetic Field Configuration}
\label{sec:Pol}

The most puzzling aspects of the GCDX is its high temperature.
Since the $kT \approx 7\,$keV plasma is
too hot to be gravitationally bound,
it requires a huge energy source
{\it without} the confinement of the plasma.
One idea to address this energetics problem is the confinement of the plasma 
by magnetic fields \citep[MFs;][]{Makishima94nhxr,99Tanuma}.
If a large-scale toroidal MF is developed and sustained, 
and the MF is strong enough for nearly energy equipartition 
with the plasma, 
the GCDX could be almost confined within the NB.
However, the large scale MF configuration was thought to be predominantly vertical,
suggesting that the magnetic confinement does not work well,
although observations of the MF in the NB have been limited to the region
in dense molecular clouds \citep{Novak00,Novak03,Chuss03}
and very thin, non-thermal radio filaments \citep{Tsuboi86,Yusef97NRTMag,Lang99}.
Recent observations have revealed that
NIR and wide-field polarimetry offers a promising tool
to trace a large-scale MF,
and a troidal configuration near the Galactic plane 
has been claimed in the GC \citep{Nishi10}.

\begin{figure}[h]
 \begin{center}
   \epsscale{.95}{
   \plotone{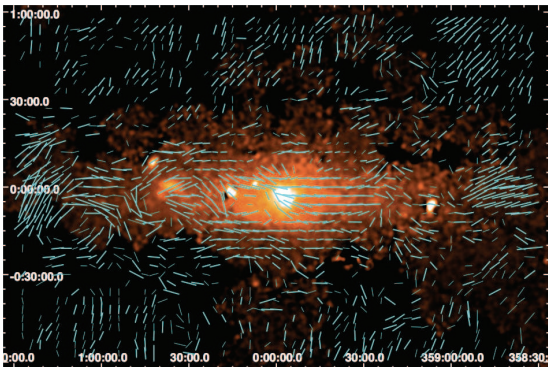}
 }
 \end{center}
    \caption{
      Polarimetry results covering $3\fdg0 \times 2\fdg0$ 
      in the Galactic coordinate,
      together with an intensity map of 6.7\,keV line emission
      \citep{Nobukawa12AIPC}.
      The cyan vectors show the inferred magnetic field direction,
      and the lengths are proportional to polarization percentage.
      The vectors are averaged in a circle of 2\farcm4 radius with a 3\farcm0 grid,
      and plotted with thick bars (detected with more than $3 \sigma$)
      and thin bars (detected with $2-3 \sigma$).
    }
   \label{fig:MagFMap}
\end{figure}

\begin{figure}[h]
 \begin{center}
   \epsscale{.80}
   \plotone{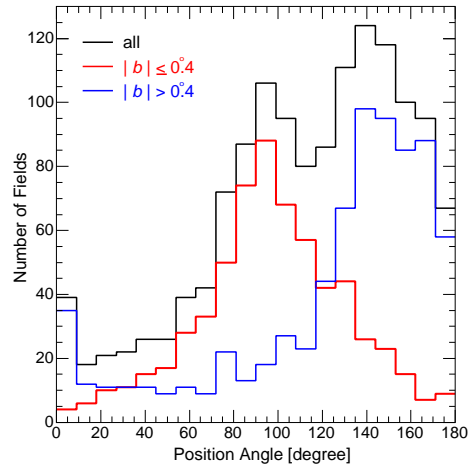}
 \end{center}
    \caption{
      Histograms of the magnetic field direction
      at $\mid b \mid \leq 0\fdg4 $ (red),
      $\mid b \mid > 0\fdg4 $ (blue), and both (black).
      The red histogram has a clear peak at the direction 
      almost parallel to the Galactic plane ($90\degr$),
      while the blue one has a peak at $\sim 150\degr$,
      almost perpendicular to the plane.
    }
   \label{fig:MagFHist}
\end{figure}

The obtained polarization map (Fig. \ref{fig:MagFMap}) suggests 
a large-scale toroidal MF configuration in the NB. 
The histogram (Fig. \ref{fig:MagFHist})
of the MF directions at $\mid b \mid < 0\fdg4$ 
has a clear peak at $90\degr$ 
which is the direction parallel to the Galactic plane.
On the other hand, at high Galactic latitude ($|b| \ga 0\fdg4$),
the fields are nearly perpendicular to the plane,
i.e., poloidal configuration.
This suggests a transition of the large-scale configuration,
and such a transition can be naturally explained by
the time evolution of MFs.
An initially predominantly poloidal, larger-scale MF is sheared out
in the azimuthal direction by the differential rotation of 
an accreting gas disk \citep{Uchida85Nat}.
The transition region, $b \sim 0\fdg3-0\fdg4$, is in good agreement 
with the scale height of the 6.7\,keV emission, $0\fdg27$ 
\citep{Uchiyama13}.

The determination of the MF strength
is still quite difficult in this region, 
but it seems to converge
to the value of $50 \la B\,\mathrm{[\mu\,G]} \la 200$ 
\citep[e.g.,][]{Ferriere11ASPC,Crocker11Heart}.
At this field strength, 
the magnetic energy density 
($\sim 0.1 - 1$\,keV\,cm$^{-3}$)
reaches nearly equipartition 
with those of diffuse hot plasma 
($\sim 0.5\,$keV\,cm$^{-3}$)
and gas turbulence
\citep[see Fig. 4 in][]{Crocker10Nat}.
This suggests that MFs provide significant pressure support
against the diffusion of the hot plasma.

If the plasma were not supported, 
it would be rushing out of the Galactic plane vertically as a galactic wind.
The escape velocity, 
typically several hundred km\,s$^{-1}$,
is smaller than the sound speed of the $7\,$keV hot plasma
of $\sim 1,400$\,km\,s$^{-1}$.
Assuming the gas flows out from the X-ray emitting region 
at the sound speed,
the escape timescale is $\sim 4 \times 10^4$\,yr \citep{Belmont05}.
This requires a huge energy input to sustain the hot plasma;
e.g., an unreasonably high 
supernova rate of $\sim 5 \times 10^{-3}$\,yr$^{-1}$
\citep[][but note that $kT \approx 7\,$keV plasma is hotter
than is observed in supernova remnants]{Uchiyama13}.
If the plasma is magnetically confined, and there is no other cooling mechanism, 
the hot plasma only cools by radiation with a timescale of 
$10^7 - 10^8$\,yr \citep{Muno04DifXGC},
several orders of magnitude longer than the escape timescale.
This would reduce the required energy input by several orders of magnitude
and thus relax the energetics problem.

There is no widely accepted mechanism
to heat the plasma to $kT \approx 7\,$keV.
If past activities of the supermassive black hole \citep{96KoyamaGCX}
heat ambient interstellar gas,
created plasma could be confined magnetically.
If a troidal MF is sustained by a differential rotation of a gaseous disk,
magnetic reconnection is a possible heating 
mechanism \citep{99Tanuma}.
Star formation activities and resultant supernovae 
are also implied to be the origin \citep{Crocker12},
while supernova-driven outflows advect materials out of the GC.
However, outflows from the GC seems to be magnetized \citep{Carretti13Nat},
hence the existence of a large-scale troidal field 
and outflows might not be contradictory.

\section{Summary}

We have used imaging and polarimetric data sets of the GC region
to investigate the origin of the GCDX.
We have constructed a stellar number density map,
and compared its longitudinal and latitudinal profiles
with those of the 6.7\,keV emission.
We have estimated that the contributions of the old stellar population
to the GCDX at the NSD and NSC are 
$\sim 50$\,\% and $\sim 20$\,\%, respectively.
Our findings support the notion that 
the GCDX is not only caused by a population of unresolved point sources 
but must also stem from a hot interstellar plasma component.
Polarimetric observations reveal
a large scale toroidal magnetic field configuration
which allows a magnetic confinement of the hot plasma.

\acknowledgements

  This work was supported by JSPS KAKENHI 
  Grant numbers 23840044, 22000005, 25707012, 
  Grant-in-Aid for the JSPS Fellows 20$\cdot$868, 
  and Excellent Young Researcher Overseas Visit Program.
  This work has made use of the IAC-STAR Synthetic CMD computation code. 
  IAC-STAR is supported and maintained by the IAC's IT Division.

\end{document}